\documentclass[fleqn,twoside]{article}
\usepackage[headings]{espcrc2}

\readRCS
$Id: espcrc2.tex,v 1.2 2004/02/24 11:22:11 spepping Exp $
\usepackage{epsfig}
\usepackage[figuresright]{rotating}

\newcommand{\AmS}{{\protect\the\textfont2
  A\kern-.1667em\lower.5ex\hbox{M}\kern-.125emS}}

\title{Design and 
Characterization of a Neutron Calibration Facility\\ for the Study of sub-keV 
Nuclear Recoils.}

\author{P.S. Barbeau\address[UC]{Department of Physics, Enrico 
Fermi Institute and Kavli Institute for Cosmological Physics\\ 
        University of Chicago, Chicago, IL 60637, USA},
        J.I. Collar\addressmark[UC]\thanks{contact author: 
	collar@uchicago.edu} and
        P.M. Whaley\address{Department of Mechanical and Nuclear 
	Engineering, \\
        Kansas State University, Manhattan, KS 66506, USA}}
       
\runtitle{}
\runauthor{}

\begin{document}

\begin{abstract}
As part of an experimental effort to demonstrate sensitivity in a 
large-mass 
detector to the ultra-low energy 
 recoils expected from coherent neutrino-nucleus elastic scattering, 
 we have designed and built a highly monochromatic 24 keV neutron 
 beam at the Kansas State University Triga Mark-II reactor. The beam 
 characteristics were chosen so as to mimic the soft recoil energies expected from reactor 
 antineutrinos in a variety of targets, allowing to understand 
 the response of dedicated detector technologies in this yet unexplored sub-keV 
 recoil 
 range. A full characterization of the beam properties (intensity, 
 monochromaticity, contaminations, beam profile) is presented, 
together with first tests of the calibration setup using proton 
recoils in organic 
scintillator.
\vspace{1pc}
\end{abstract}

\maketitle

\section{INTRODUCTION}

The scattering of low-energy neutrinos off 
nuclei ($E_{\nu}\!<$ few tens of MeV, e.g., reactor 
$\bar{\nu} s$) via the neutral current 
remains undetected thirty years after its first description \cite{freedman}. 
The small momenta exchanged at 
these neutrino energies result in  
the entire nucleus being probed, giving rise to a large coherent 
enhancement to the cross 
section, roughly proportional to neutron number squared \cite{drukier}.
The resulting cross section is very large
compared to other low-energy neutrino interaction channels. 
This process 
is quantum-mechanically possible due to the indistinguishability of initial and final 
target states, having no charged-current equivalent.
Using this mode of interaction, it would be possible to imagine
smallish neutrino detectors: in some experimental conditions 
the expected rates can be as high as several 
hundred recoils/kg day \cite{drukier,meyannis}, by no means a ``rare-event'' situation. 
However, the recoil energy transferred to the target is of a few keV at most for 
the lightest nuclei, with only 
a few percent of it going into readily measurable channels (ionization or 
scintillation). As a result of this, no realistic detector technology 
has been available with the combination of mass and energy threshold 
capable of a first measurement of this intriguing mode of neutrino 
interaction. 

This paper is part one of two: here
the steps taken to demonstrate sensitivity to sub-keV recoils in 
a variety of detecting media by means of a low-energy highly 
monochromatic neutron beam are delineated. In a companion paper \cite{ibid}
a new detector technology (a large-mass modified electrode p-type 
germanium diode) 
finally able to exploit this mode of neutrino 
interaction is described, as well as the expected performance and applications  
of this detector 
in fundamental and 
applied neutrino physics, and in other 
areas (e.g., dark matter searches).

\section{A CALIBRATION FACILITY FOR COHERENT NEUTRINO DETECTORS, MINUS 
THE NEUTRINOS}

Devising a method able to accurately determine the response of a  detector 
to sub-keV recoils is almost as difficult as developing 
a technology sensitive to these. The first coherent 
neutrino detector will be entering a low-energy {\it terra 
incognita}, never explored before with devices of any significant 
mass. It would be foolish to expose a capable detector to a high-flux of 
reactor antineutrinos to then claim that a
low-energy excess dangerously neighboring the electronic noise 
originates in 
the sought recoils: a very precise understanding of quenching factors 
(i.e., the fraction of the recoil energy available for detection in 
the form of ionization, scintillation, etc.) in this energy region is 
a must, prior to exposure to a neutrino source. Towards this end at 
least two methods have been put forward. One relies on the recoiling 
daughters (of up to a few hundreds of eV) that accompany high-energy 
gamma emission following thermal neutron 
absorption \cite{ieee,ge}. The disadvantage of this 
approach is a very limited choice, if any, of intense enough gamma-emitting
transitions 
leading to energetic enough recoils in
any given detector material. An aggravating factor 
is that not all decays produce predictable energy depositions, due to prompt 
secondary gamma emission while the recoil is still slowing down \cite{ge}. A 
better approach \cite{meyannis} is to rely instead on the existence of 
narrow deep interference ``dips'' in the neutron cross-sections in a handful of 
isotopes (\frenchspacing{Fig. 1}): this can be exploited to build transmission filters 
in nuclear reactor facilities that allow  neutrons only in those 
precise energy bands 
to percolate \cite{nbeams}.
In the case of an iron filter, 
the 24 keV ($\pm$2 keV FWHM) neutron emission is ideal to mimic the recoil 
energies expected from reactor antineutrinos 
(\frenchspacing{Fig. 2}) \cite{meyannis}. Iron is a good 
gamma shielding material and well-designed filters of this type 
generally exhibit a very low contamination with these or with neutrons in 
other energy ranges \cite{nbeams}. An additional aluminum post-filter
has the effect of reducing the overall neutron flux, but 
preferentially at larger than 24 keV energies, i.e., results in a 
higher beam purity (\frenchspacing{Fig. 1}). A thin ($\sim$1 cm) 
removable titanium post-filter can be 
used to scatter away the 24 keV component without affecting other 
beam components (i.e., the 
backgrounds), a result of a resonance in the Ti neutron cross-section,
centered precisely around this energy (\frenchspacing{Fig. 1}). This provides a unique opportunity to perform ``signal 
on'' and ``signal off'' runs, allowing the experimenter to make sure 
that tiny 
recoil-like signals are 
indeed originating from the 24 keV component and not in any beam 
contamination, no matter how secondary. The beneficial effect of this on the 
credibility of the results cannot be overemphasized.

\begin{figure}[tbp]
\epsfxsize = 8.4 cm
\begin{center}
\epsfbox{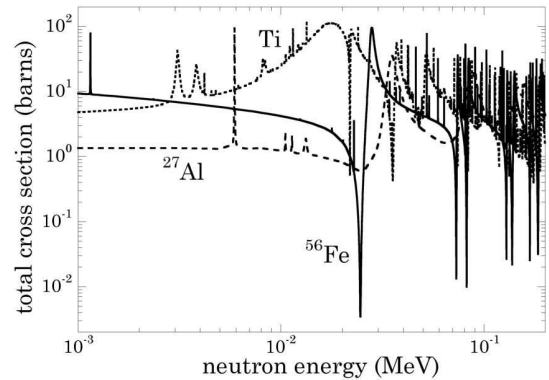}
\end{center}
\caption{{\scriptsize Neutron cross-sections of relevance in the design of a 
24 keV iron filter beam. Aluminum is considered an ideal post-filter 
material, able to preferentially suppress surviving neutron energies 
$>$24 keV. Titanium can be used to ``switch off'' the main 24 keV 
component, while leaving all other neutron energies and any gamma rays 
essentially unaltered.}} 
\end{figure}

\begin{figure}[tbp]
\epsfxsize = 8.2 cm
\begin{center}
\epsfbox{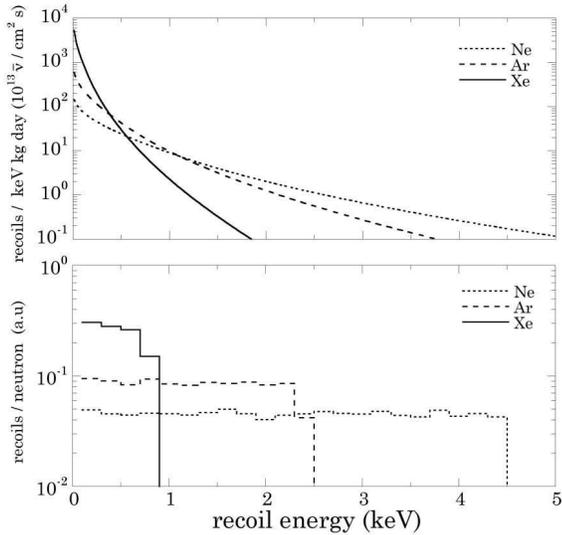}
\end{center}
\caption{{\scriptsize Top: Spectrum of recoil energies 
expected from reactor antineutrinos in several target materials. 
Bottom: Recoil spectrum similar in energy span, from exposure to a 24 keV 
neutron beam (MCNP-Polimi simulation \protect\cite{polimi}).}} 
\end{figure}

\subsection{Design of a 24 keV neutron beam}\label{sec:¥}

\begin{figure}[tbp]
\epsfxsize = 7.4 cm
\begin{center}
\epsfbox{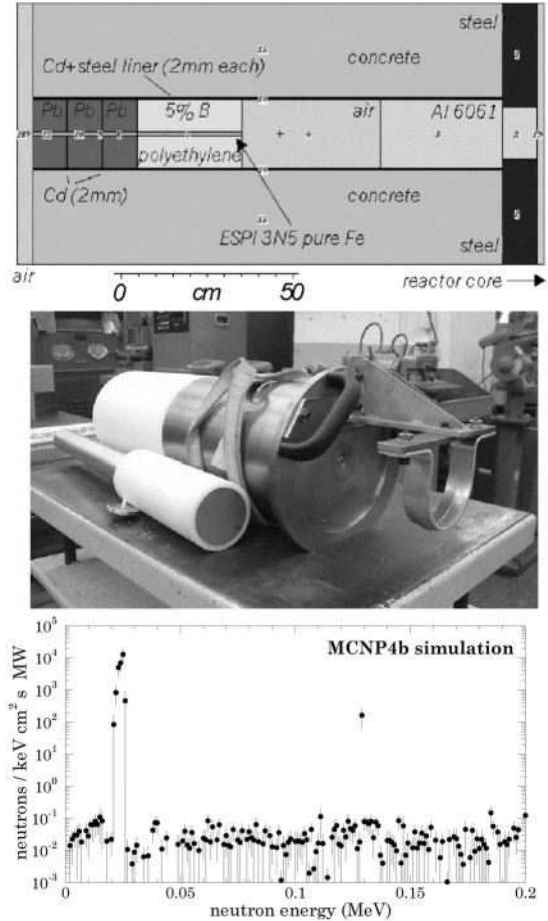}
\end{center}
\caption{{\scriptsize Top: Schematic design of the UC/KSU filter. 
The Fe rod was machined with a tapered cross-section to avoid streaming of 
radiation. The Al spectrum shifter was removed in the 
final configuration. Middle: Completed filter, with Al/Ti post-filter holder 
visible in the foreground. A laser alignment tool can be 
screwed onto the Fe rod, allowing precise detector placement.
Bottom: Predicted beam composition.}} 
\end{figure}

A present penury of iron filter reactor beams in the US lead 
to the design and 
construction of a dedicated, easily-removable filter at the University of Chicago, for installation 
in the Kansas State University TRIGA Mark-II experimental reactor. 
MCNP4b \cite{mcnp} simulations using available spectral measurements of 
unfiltered tangential beam neutron emissions as the input, revealed that a narrow 
(2.5 cm diameter, 60 cm long) filter extremely low in non-Fe
contaminants \cite{espi} would produce an  
equivalent beam intensity and purity to a larger diameter, lower 
quality iron rod, which is a more common approach \cite{nbeams} 
(``beam purity'' is the fraction of the total neutron flux falling 
under the 24 keV peak).
\frenchspacing{Fig. 3} displays an schematic of the final 
design. These simulations (\frenchspacing{Fig. 3}, bottom) demonstrated that a high-enough 
24 keV flux and purity should be available even from a
low-power experimental reactor (240 kW max., with a current upgrade 
to 1.25 MW pending). During initial testing, a gamma background
ignored in the simulations
was observed (50 mRem/hr on contact), streaming from the concrete in the 
periphery of the beam, originating in moderation and capture of neutrons once-scattered in the 
back of the filter. This background was controlled with the addition 
of 50 cm of high-density concrete and Pb shielding around the beam exit.

\subsection{Beam Characterization}\label{sec:¥}
\subsubsection{Neutron flux and angular divergence}\label{sec:¥}

Neutron flux measurements using a ``Benjamin'' spherical proton-recoil 
spectrometer \cite{benj1} placed both along the beam axis and off-axis 
revealed a measurable angular divergence. 
An equivalent point source, derived from axial measurements,
was seen to reside -36.1$\pm$4.8 cm into the filter (taking the 
exit point as the origin of coordinates), in good agreement with 
a geometric neutron optics estimate (z=-35.9 cm). Based on off-axis measurements, the beam 
lateral spread at the z=+110 cm location (the chosen detector 
calibration point)
was observed to span $\sim$6 cm FWHM. This was confirmed by careful 
mapping of this 
spread with a small (1 c.c.) $^{6}$LiI(Eu) scintillator \cite{proteus}, 
95\% enriched in $^{6}$Li, surrounded by 0.5 mm of Cd to reduce 
response to thermal neutrons. The 2-d mapping resulting from 
measurements over a 10$\times$10 grid of positions, displayed in 
\frenchspacing{Fig. 4}, is an accurate 
description of the local
neutron flux at the detector irradiation point. The 24 
keV neutron flux component was extracted from the counting rate difference between 
Ti-on and Ti-off operation, with a detector efficiency (reaction rate) derived from 
MCNP4b simulations.
The maximum beam intensity at the filter exit 
was measured using the proton-recoil spectrometer at 7.9$\times$10$^{4}$ ($\pm$10\% stat. 
$\pm$15\% sys.) n cm$^{-2}$ 
s$^{-1}$ MW$^{-1}$, a typical value for Fe filter designs \cite{nbeams}.

\begin{figure}[tbp]
\epsfxsize = 7.4 cm
\begin{center}
\epsfbox{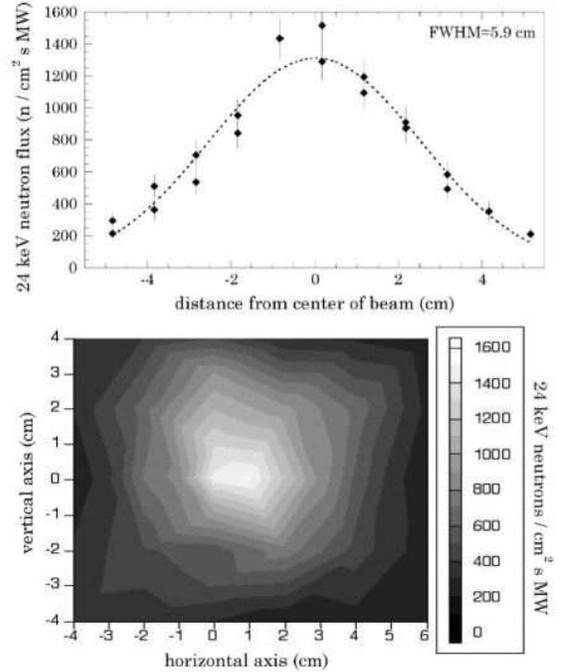}
\end{center}
\caption{{\scriptsize Top: Beam profile characterization at the detector irradiation 
point, 110 cm in front of the beam exit, using a small enriched 
$^{6}$LiI(Eu) scintillator.
Bottom: A 2-d image of the same, revealing a small lateral 
misalignment of $\sim$0.5 cm, a measure of the present ability to 
center 
detectors under test with the beam, using a laser tool.}} 
\end{figure}

\begin{figure}[tbp]
\epsfxsize = 7.4 cm
\begin{center}
\epsfbox{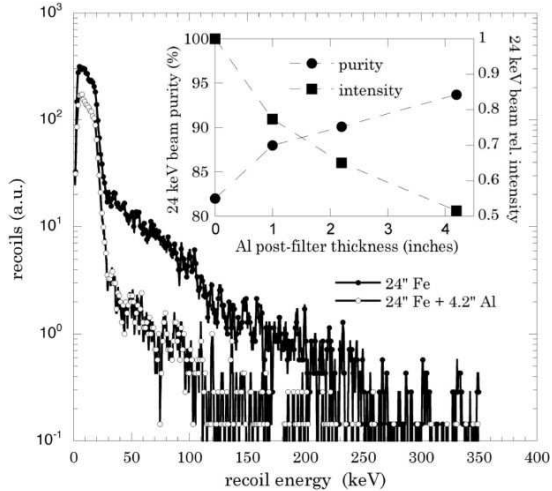}
\end{center}
\caption{{\scriptsize Effect of an Al post-filter in increasing the purity of a 24 
keV Fe neutron filter (``purity'' defined as the percent of  
neutrons emitted falling under the 24 keV peak). The spectra (same 
exposure) are measured
using a spherical (``Benjamin'') hydrogen-filled proton-recoil 
spectrometer. Recoil energies of up to the incident neutron kinetic energy 
are possible in this type of detector.}} 
\end{figure}

\begin{figure}[tbp]
\epsfxsize = 7.4 cm
\begin{center}
\epsfbox{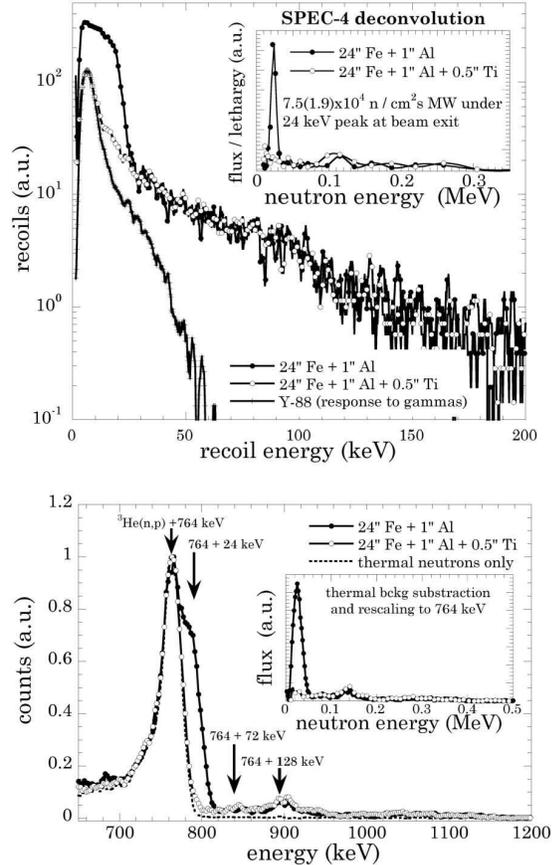}
\end{center}
\caption{{\scriptsize Top: Dramatic effect of a thin Ti post-filter in removing the main 
24 keV beam component while allowing contaminations to pass through (see 
text). Ti filter on and off spectra correspond to the same exposure. Bottom: A 
lower in resolution (yet similar in features) energy spectrum revealed 
by a 
$^{3}$He ionization chamber, used as a rudimentary spectrometer. The normalized 
response of the detector to a pure thermal neutron field is shown for 
comparison.}} 
\end{figure}

\begin{figure}[tbp]
\epsfxsize = 7.4 cm
\begin{center}
\epsfbox{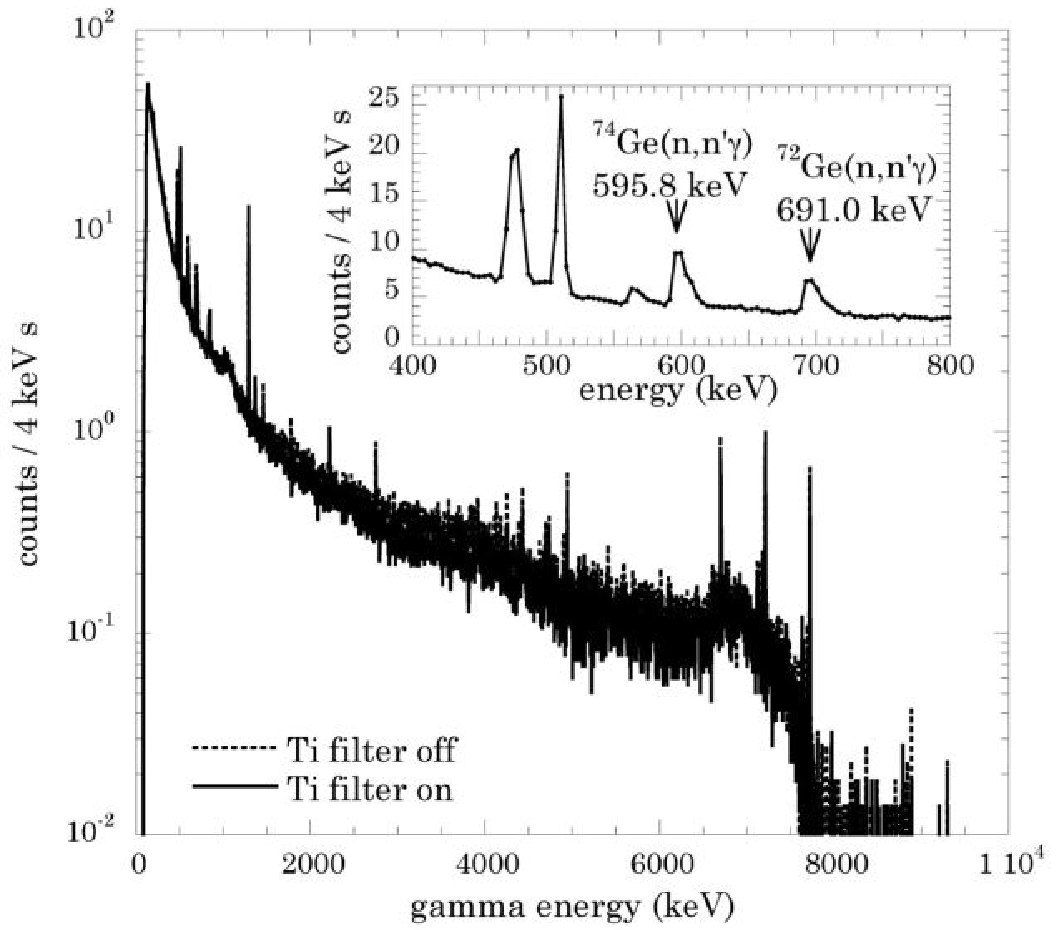}
\end{center}
\caption{{\scriptsize Negligible effect of Ti 
post-filter on the very small gamma beam contamination, as measured by a HPGe detector. 
The inset displays characteristic asymmetric peaks from inelastic neutron 
scattering, used to set limits on beam contamination by fast neutrons
(see text). Reactor power was at the present maximum of 
240 kW during these measurements.}} 
\end{figure}

\subsubsection{Monochromaticity and beam contaminations}\label{sec:¥}

Additional measurements of axial beam intensity, composition, effect 
of filters and radial divergence were 
performed using the proton-recoil 
spectrometer, a $^{3}$He ionization chamber and a HPGe gamma detector.  
The first allows to perform high-resolution neutron spectroscopy by 
deconvolution (unfolding) of its response, down to $\sim$1 keV neutron 
energies \cite{benj2}. 
\frenchspacing{Figs. 
5-7}
encapsulate these results (datapoints in \frenchspacing{Fig. 
5} and top \frenchspacing{Fig. 
6} are modestly smoothed to facilitate comparison and deconvolution). 
The  beam purity obtained in the presence 
of a 2.5 cm Al post-filter \cite{alu} is $\sim$88 \% (\frenchspacing{Fig. 5}), 
as measured at z=+110 cm. The marked decrease in $>$24 keV recoils by the 
addition of the Al post-filter is clearly observed in the figure, at the expense of a  
comparatively much smaller decrease in the 24 keV component. This 
effect can 
be understood by inspection of the neutron cross-sections in \frenchspacing{Fig. 1}.

\frenchspacing{Fig. 6} (top) displays the dramatic 
effect of the addition of a 1.25 cm-thick Ti post-filter: the 24 keV 
peak intensity is squelched by a factor $\sim$ 20, in agreement with 
simulations, while the comparatively  
smaller flux of neutrons at all other 
energies remains essentially unaffected, providing a way to ``switch off'' 
exclusively the source of the soft recoils of interest. The response 
of the proton-recoil spectrometer to an intense (3 mCi)
$^{88}$Y high-energy gamma source is 
normalized in the figure
to the  maximum of the  Ti filter spectrum, illustrating the 
marginal sensitivity to gammas in this type of detector, which allows
to perform low-energy neutron measurements \protect \cite{benj2}. The inset is the
deconvolution of the proton recoil spectrum to generate a 
neutron energy spectrum, using the SPEC-4 
code \cite{spec4}. It clearly demonstrates the good monochromaticity 
achieved and the extinction of exclusively the 24 keV peak in the 
presence of Ti.  This was also confirmed by the 
$^{3}$He measurements (Fig. 6, bottom), using this type of counter as a 
rudimentary spectrometer. The non-proportionality of this 
last detector results in a much diminished energy resolution in the 
unfolded spectra, but identical beam components are revealed. The small higher 
energy components (72 keV, 128 keV, etc.) discernible in both 
proton-recoil and $^{3}$He unfolded spectra are typical of Fe filters 
\protect \cite{nbeams,ahl} and the 
result of second-order passing bands in the Fe cross-sections 
(\frenchspacing{Fig. 
1, 3 bottom}), not entirely canceled by the Al post-filter. 

The 
effect of the thin Ti post-filter on gamma backgrounds is also negligible, 
leading to a mere 6\% reduction in overall gamma flux (\frenchspacing{Fig. 7}). 
The presence of several characteristically asymmetric peaks in the 
HPGe spectrum, arising from inelastic neutron scattering, 
(\frenchspacing{Fig. 7}, inset) allows to impose further limits on the neutron 
component at energies $>$691 keV \cite{gen} (less than 100 n cm$^{-2}$ 
s$^{-1}$ MW$^{-1}$ at z=+110 cm). Similarly, a thermal absorption peak in Ge at 139.9 keV 
\cite{gen} allows 
to set the thermal neutron flux in the irradiation point at $\sim$30 n cm$^{-2}$ 
s$^{-1}$ MW$^{-1}$, a very low value characteristic of Fe filters 
\cite{nbeams}. The gamma dose at z=+110 cm was found to be a 
very modest 0.3 mRem/hr at full reactor power (240 kW).

\subsection{Tests of the calibration setup}\label{sec:¥}

\begin{figure}[tbp]
\epsfxsize = 7.4 cm
\begin{center}
\epsfbox{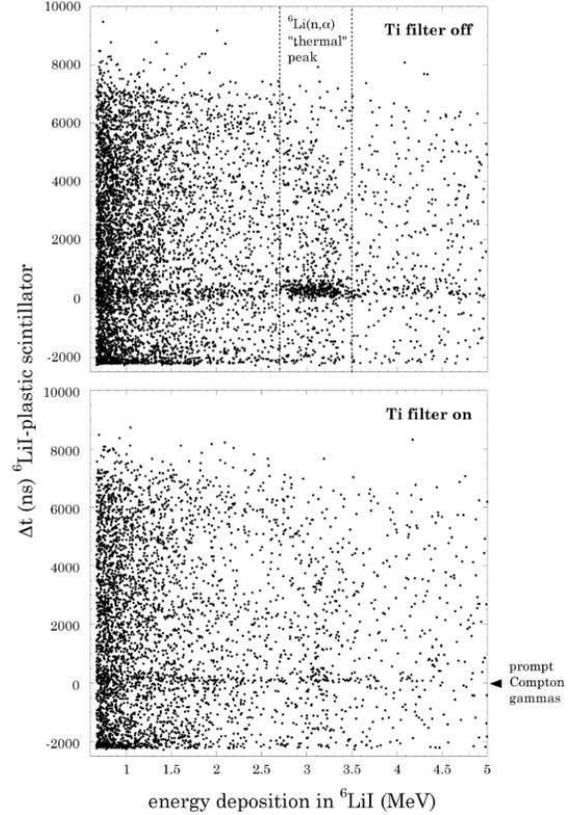}
\end{center}
\caption{{\scriptsize  Measured time-difference between low-energy hydrogen recoils in BC-404 
plastic scintillator and detection of the scattered neutron via 
$^{6}$Li(n,$\alpha$). Correlated true-coincidences induced by the 24 keV 
component of the beam can be seen to vanish in the presence of 
the Ti post-filter. Both figures correspond to the same exposure.}} 
\end{figure}

\begin{figure}[tbp]
\epsfxsize = 7.4 cm
\begin{center}
\epsfbox{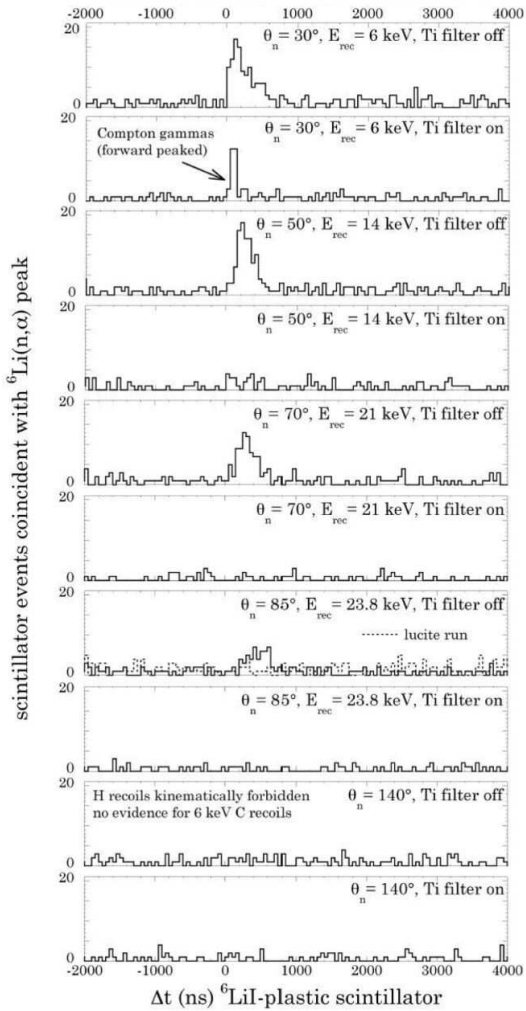}
\end{center}
\caption{{\scriptsize  Total of all BC-404 scintillator 
measurements (see text). An optimal ability 
to positively
identify sub-keV recoils in detectors capable of coherent 
neutrino detection has been demonstrated in these tests of the UC/KSU 
beam and setup.}} 
\end{figure}

\begin{figure}[tbp]
\epsfxsize = 7.4 cm
\begin{center}
\epsfbox{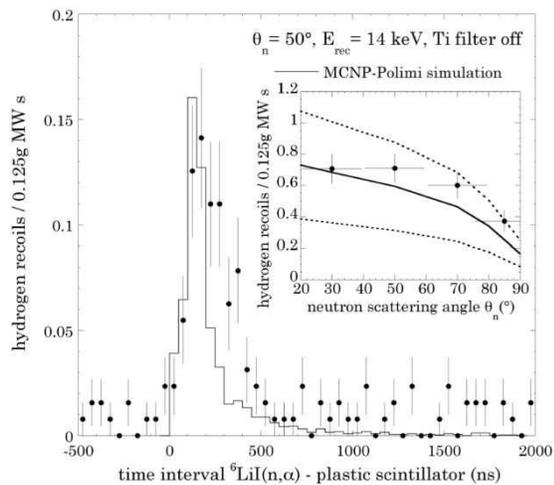}
\end{center}
\caption{{\scriptsize  Comparison of time differences between BC-404 
hydrogen recoil
and $^{6}$LiI(Eu) neutron capture
signals with a MCNP-Polimi \protect\cite{polimi}
simulation devoid of free parameters. The effect of neutron time-of-flight between both 
detectors and straggling 
within the $^{6}$LiI(Eu) scintillator prior to capture is evident. 
The simulation takes into account the finite timing capabilities of 
the data acquisition system. Inset: measured rate 
of true coincidences between
hydrogen recoils in BC-404 and the $^{6}$Li(n,$\alpha$) 
scattered neutron capture signal, as a function of $\theta_{n}$. The 
solid line corresponds to Monte Carlo expectations. 
Dotted lines represent a conservative one sigma uncertainty in these, dominated by 
a 15\% estimated error assigned to each of two Monte Carlo simulations involved.}} 
\end{figure}

\begin{figure}[tbp]
\epsfxsize = 7.4 cm
\begin{center}
\epsfbox{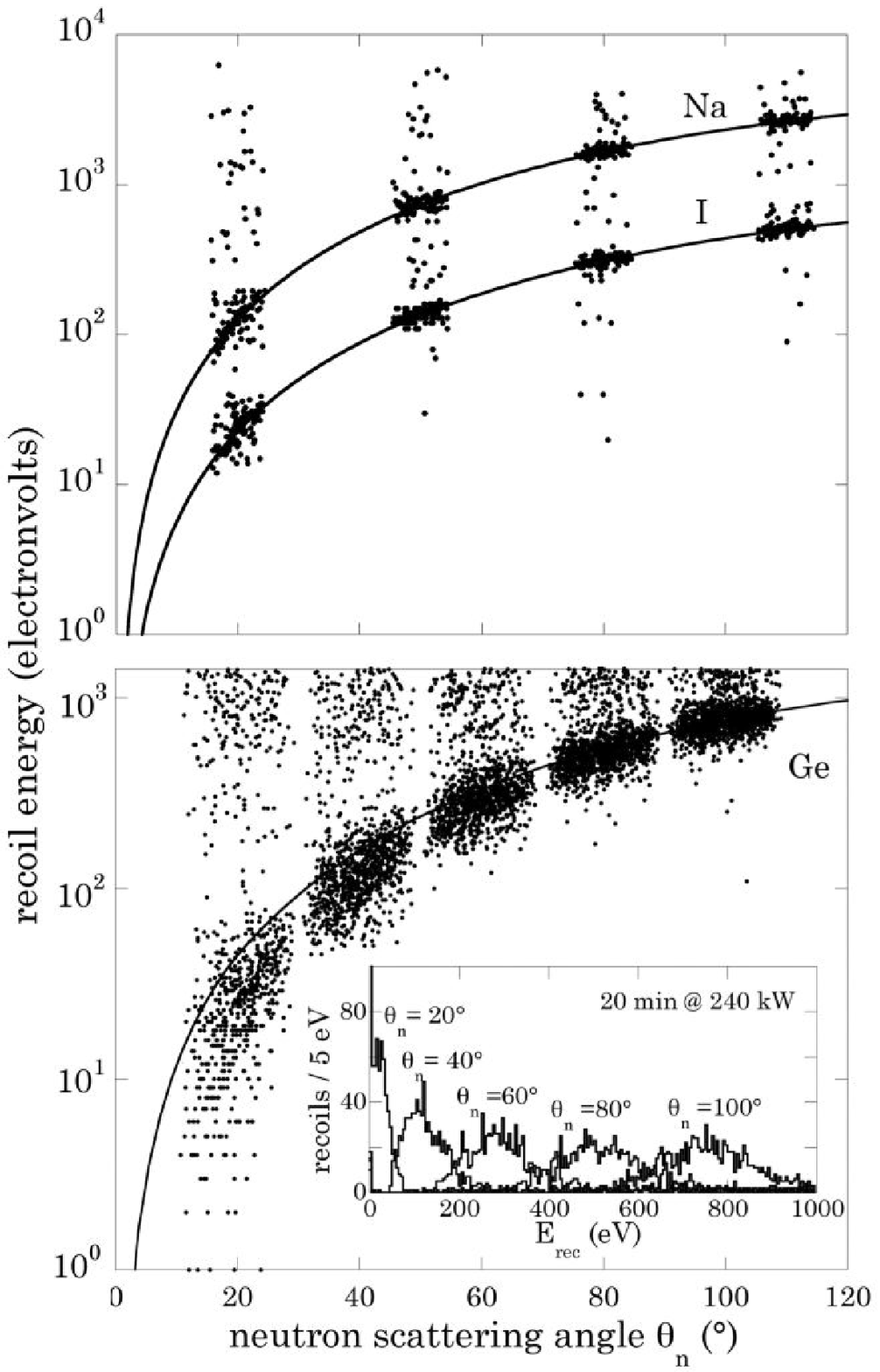}
\end{center}
\caption{{\scriptsize  MCNP-Polimi simulated recoil rates and energies in 
the UC/KSU beam. Solid lines are the 
kinematic expectations. Top: small (1 c.c.) NaI scintillator. Bottom: 
500 g HPGe detector (multiple scattering in 
this larger crystal leads to a small fraction of events with 
energy depositions of up to a few keV). The inset shows the signal 
rates expected in coincidence with the $^{6}$LiI(Eu) detector described 
in the text. This modified-electrode HPGe detector and the results of 
its calibration in this facility are the subject of 
a companion paper \protect\cite{ibid}.}} 
\end{figure}

It is clear from \frenchspacing{Fig. 2} that a 24 keV neutron beam 
provides a unique 
opportunity to produce sub-keV nuclear recoils on any target, much like those expected 
from reactor antineutrinos. The recoil energy can be selected at will 
by tagging the neutron scattering angle, $\theta_{n}$, with a suitable auxiliary 
neutron detector. The highest possible efficiency in the detection of 
scattered neutrons is needed if a low power experimental reactor is to 
be used: Monte Carlo studies of the response of a number of possibilities
lead to the choice of a large (5 cm diam. x 7.5 cm long) enriched $^{6}$LiI(Eu) crystal as optimal 
for scattered neutrons in the few keV range. While $^{6}$Li is 
typically used for thermal neutron detection, a large crystal of this size 
allows for efficient moderation of intermediate neutron energies within the 
crystal itself, followed by capture: at 97\% $^{6}$Li enrichment, 
29\% of collimated 24 keV neutrons entering 
through the front 
face of this crystal are expected to be captured.
A second important advantage from this type of detector is the 
appearance of the $^{6}$Li(n,$\alpha$) absorption ``thermal peak'' at 
3.1 MeV electron equivalent, an energy high enough 
for most gamma backgrounds to fall below it, allowing for an efficient 
environmental gamma background reduction. The rate under this ``thermal'' peak, 
revealing of the arrival of a scattered neutron, is 
 $\sim$ 100 Hz at full reactor power (240 kW), with just a thin layer of borated silicone as (thermal 
neutron) shielding around the crystal. This is a comfortably low figure that 
ensures an absence of spurious coincidences between the detector 
under test and the $^{6}$LiI(Eu)  crystal \cite{nota}.

A repetition of a historic 
experiment using a similar beam was performed 
to test the full calibration system \cite{ahl}. In the original work, an 
unexpectedly-high light yield from sub-keV hydrogen recoils was 
measured in plastic 
scintillator, demonstrating the ability to 
look for slow magnetic monopoles with the MACRO detector \cite{macro}. 
In the present 
remake, a sample of high yield BC-404 plastic 
scintillator small enough  
(0.5 c.c) to ensure avoiding multiple scattering  
was used as the target detector in front of the beam. 
Because of the small hydrogen recoil energies involved (24 keV maximum), low 
light yield of organic scintillators and 
quenching factors of O(10)\% \cite{ahl} the signals expected in coincidence 
with the $^{6}$Li(n,$\alpha$) peak involve no more than a few 
photoelectrons, even for a well-matched photomultiplier. Nevertheless a 
time-coincident excess of events is clearly observed in the absence of 
the Ti post-filter, to vanish in its presence (\frenchspacing{Fig. 
8}). \frenchspacing{Fig. 9}
displays the total of the measurements performed at different values 
of $\theta_{n}$. Several interesting (yet
expected) features are present in the data. For instance, the 
appearance of prompt coincidences due to forward-peaked Compton 
scattering of gammas only at small $\theta_{n}$. The neutron time-of-flight 
between both detectors is also visible ($\sim$70 ns for 15 cm @ 24 keV), as well as the delay from neutron 
straggling within the $^{6}$LiI(Eu) crystal, in excellent agreement with 
MCNP-Polimi simulations \cite{polimi} (\frenchspacing{Fig. 10}). 
Finally, as expected, no hydrogen recoils 
were observed at kinematically-forbidden $\theta_{n}\!\!>90^{\circ}$. Carbon 
recoils as low as few keV might nevertheless produce enough 
scintillation light to be
detectable at those angles \cite{new1}, 
but the much lower recoil rate expected for these and the short exposure 
during these first tests (1 hr per angle) did not lead to an 
observable occurrence. As originally performed in \cite{ahl}, 
the BC404 scintillator was replaced during a run by an 
identical sample of lucite to test that coincidences occur from 
actual recoils in the BC404
and not from Cerenkov light in the photomultiplier glass envelope: no obvious 
coincidences attributed to the envelope were observed 
(\frenchspacing{Fig. 9}).

\section{CONCLUSIONS}

As evidenced by \frenchspacing{Figs. 8-10}, the setup, beam purity and control of 
backgrounds achieved allows to unequivocally identify events in a detector 
under test that originate in ultra-low energy nuclear recoils (well below 1 keV 
for targets heavier than hydrogen), with a convenient ability 
to select their energy. This is achieved in excellent signal-to-noise conditions, by 
requesting coincidences with the $^{6}$Li(n,$\alpha$) peak. The 
additional
ability to ``switch off'' these low-energy recoils by means of the Ti 
post-filter, while allowing the scarce
backgrounds to remain unaffected, provides a  
convenient and convincing workbench for calibration of coherent neutrino detector 
technologies. \frenchspacing{Fig. 11} displays the expected range of 
recoil energies and signal rates that the beam and calibration setup 
can provide, for two different target materials. 

As discussed in the companion paper \cite{ibid}, at least one detector 
technology has been developed
with all the characteristics needed to attempt an exciting first reactor measurement 
of the coherent neutrino scattering cross section. The tools are 
already in place for the 
much needed preliminary step, a careful calibration using soft recoils similar to those  
from reactor antineutrinos.

\section{ACKNOWLEDGEMENTS}
We are indebted to the instrument design makers at EFI, E. Mendoza, 
R. Metz, D. Plitt and G. Ward, for the dedication 
and finesse with which the filter was constructed. Also to A. Cebula, 
E. Cullens, A. Meyer, L. Retzlaff, C.J. Solomon, Troy 
Unruh, R. Van Fange, J. Van Meter
and the rest of the KSU reactor operators for their generosity and patience in running for 
many long hours, and their dexterity in keeping reactor poisoning at 
bay. This work is supported by NSF CAREER award PHY-0239812, NNSA 
grant DE-FG52-0-5NA25686 and in part by 
the Kavli Institute for Cosmological Physics through grant NSF 
PHY-0114422 and a Research Innovation Award 
No. RI0917 (Research Corporation).


\begin{thebibliography}{9}
\bibitem{freedman} D. Z. Freedman {\it et al.}, Annu. Rev. Nucl. Sci. {\bf 
27} (1977) 167.
\bibitem{drukier} A. Drukier and L. Stodolsky, Phys. Rev. {\bf D30} (1984) 2295.
\bibitem{meyannis}J.I. Collar and Y. Giomataris, Nucl. Instr. Meth. {\bf 
A471} (2001)
254.
\bibitem{ibid}P.S. Barbeau, J.I. Collar and O. Tench,
{\it ``Large-Mass Ultra-Low Noise 
Germanium Detectors: Performance and Applications in Neutrino and 
Astroparticle Physics''}, submitted to Phys. 
Rev. C, available from {\tt 
http://cfcp.uchicago.edu/$\sim$collar/co2.pdf}
\bibitem{ieee}P.S. Barbeau {\it et al.}, 
IEEE Trans. Nucl. Sci. {\bf 50} (2003) 1285.
\bibitem{ge}K. W. Jones and H. W. Kraner, Phys. Rev. {\bf A 11} (1975) 
1347.
\bibitem{nbeams} R. C. Block and R. M. Brugger in ``Neutron Sources for 
Basic Physics and Applications'', A. Michandon {\it et al.} eds., Pergamon, 
NY, 1983; F.Y. Tsang and R.M. Brugger, Nucl. Instr. Meth. {\bf 134} (1976) 441;
O. Aizawa {\it et al.}, J. Nuc. Sci. Tech. {\bf 20} (1983) 354; A.V. 
Murzin  {\it et al.}, Sov. At. Energy {\bf 67} (1989) 699; S.V. Musolino 
 {\it et al.}, Med. Phys. {\bf 18} (1991) 806; C.A. Perks {\it et al.}, 
Rad. Prot. Dosim. {\bf 15} (1986) 31.
\bibitem{polimi}S.A. Pozzi {\it et al.}, Nucl. Instr.  Meth. {\bf 
A513}, 550  (2003).
\bibitem{mcnp}{\it MCNP, A General Monte Carlo N-Particle Transport 
Code}, J.F. Briesmeister ed., Los Alamos National Laboratory, 
LA-12625-M, 1993. 
\bibitem{espi}3N5 pure iron (ESPI, Ashland, Oregon 97520, USA).
\bibitem{benj1}Model 27044 (LND, Oceanside, New York 11572, USA).
\bibitem{proteus}Proteus Inc., Chagrin Falls, Ohio 44022, USA.
\bibitem{alu}Alloy 1100, commercially pure (99\%) aluminum.
\bibitem{benj2}W.H. Miller, Nucl. Instr. Meth. {\bf A279} (1989) 546.
\bibitem{spec4}Oak Ridge National Laboratory, RSICC peripheral 
shielding routine collection, {\it SPEC-4, Calculated Recoil Proton Energy Distributions
from Monoenergetic and Continuous Spectrum Neutrons.}
\bibitem{nota}The original $^{6}$LiI crystal used for these 
measurements had been salvaged and 
reencapsulated from a 
larger, partially-hydrated and fractured unit, exhibiting a degraded 
resolution. A brand new crystal, 5 cm in diameter and 1.5 cm long has 
been developed by Proteus Inc. At the expense of a loss in 24 keV 
neutron efficiency by a factor $\sim3$, the background rate under the thermal 
peak is reduced by a factor $\sim20$ in this unit, greatly improving 
the capabilities of the setup.
\bibitem{ahl}D.J. Ficenec {\it et al.}, Phys. Rev.  {\bf D36} (1987) 
311; S.P. Ahlen {\it et al.}, Phys. Rev.  Lett. {\bf 55} (1985) 
181.
\bibitem{gen}R. Wordel {\it et al.}, Nucl. Instr. Meth. {\bf A369} 
(1996) 557.
\bibitem{macro}M. Ambrosio {\it et al.}, Eur. Phys. J. {\bf C25} (2002) 511.
\bibitem{new1}J. Hong {\it et al.}, Astropart. Phys. {\bf 16 }(2002) 333.
\end{thebibliography}
\end{document}